# Development of High Granulated Straw Chambers of Large Sizes

V.Davkov[1], K.Davkov[1], V.V.Myalkovskiy[1], L.Naumann[2],
V.D.Peshekhonov[1], A.A.Savenkov[1], K.S.Viryasov[1], I.A.Zhukov[1]

[1]) Joint Institute of Nuclear Research, Dubna, Russia
[2]) FZ Rossendorf, Dresden, Germany, Institut f. Kern- und Hadronenphysik

We have developed the baseline design for the straw drift tube tracking detectors for high rate environment application. The low-mass inner straw elements and the technology of the multianode straws assembly was devised and checked. The prototype chamber was constructed and studied the granularity of similar chambers can be reduced to one $cm^2$.

## 1. Introduction

Drift chambers on the basis of thin-walled drift tubes (straw) are widely used as track detectors especially in the high rate environment [1,2,3,4]. The transition radiation tracker of the ATLAS Inner Detector [2] is good example of the above applying the straw with the diameter of 4mm and length of ~40cm and 150cm for End-Cap and Barrel detectors, correspondingly. The tracker of the spectrometer COMPASS uses the straw with the diameter of ~9.6mm for periphery of the chambers and 6mm diameter - for the central area, the sensitive length of the straw is to 325cm. Granularity of the central area of the COMPASS chambers is equal to ~192$cm^2$ and ~76$cm^2$ for the small part of the area near the beam hole in them. Each straw is read out from one end and the occupancy in each tube below 2% at maximum beam rates for the intensities about $2 \cdot 10^8$ muons and up to $4 \cdot 10^7$ hadrons per 5.1 s spill for muon and hadron beams, respectively. The expected rate for the Barrel TRT straws is about 17 MHz at the luminosity $L=10^{34} cm^{-2} c^{-1}$. To reduce the straw occupancy by 2 times, the Barrel straws consist of two electrically isolated anode parts [2, 5]. The read-out is performed independently from each part of the anode from two ends for each straw.

In the set-up CBM GSI under construction the expected rate is about 140 kHz/$cm^2$ for Au-Au collisions at 25AGeV with the reaction rate of 10MHz [6]. To provide a small value of the occupancy for the track detectors of large area, their granularity should be several $cm^2$. Using the straw chambers with the anode readout will result in strong growing of this value in case of long straws. For the case when the detector is set of small modules with the short straws there will be not only a low ratio of the sensitive area to the overall detector dimensions but also an increase of its average radiation thickness because of the external module frames.

It has been earlier shown a principle opportunity of constructing straw with three electrically isolated parts of its anode and readout for the both straw ends from its end-plugs, and for the central anode segment - through the straw wall [5]. Construction of the big length straw with a multi segmented anode providing independent readout from the electrically independent segments, allows one to obtain the necessary occupancy over all the area of the large-size detector by optimizing the segmented lengths.

The development of the low-mass readout links for the inner anode segments, using new technologies, should result in preserving a minimal value of the straw detector radiation thickness in comparison with well-known coordinate detectors. Highly granulated coordinate straw detectors of large area having small radiation thickness and high radiation hardness, should be interesting for the CBM GSI, upgraded of the LHC detectors and construction of detectors for ILC.

Below there are some results of the first prototype made at JINR with multi segmented anodes for the straws 4mm in diameter.

## 2. Concept of the straw with a multisegmented anode

To construct two segmented anodes, a stand has been developed at JINR as well as methods of their manufacturing [7]. Two anode wires separated from each other by an isolated gap, are fused into a glass capillary tube 6 or 7mm length, 0.25mm in outer diameter and 0.1mm of inner diameter (Fig.1). The mass of the boronsilikate capillary tube is 0.094mg per 1mm of its length.

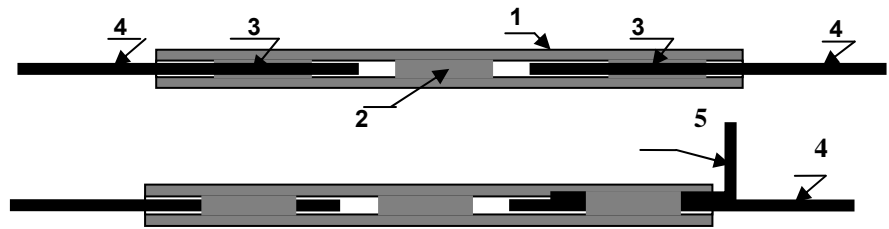

Fig.1. Capillary tube - (1), isolation gap - (2), (3)- molten location of wire. Anode wire (4) and additional contact wire (5) are the gilded tungsten wires in 30μm of diameter

The bottom part of figure shows an opportunity to fuse the additional wire and anode wire together in the capillary tube, for the galvanic contact to this segment for the readout.

A concept of manufacturing straws having a segmented anode is demonstrated with an example of the straw with three anode segments (Fig.2). A multi segmented anode assembled beforehand, is put into the straw and fixed at the ends. The readout from the end anode segments is carried out from the straw end plugs. The information from the middle segment is read out over the contact wire going through the spacer supporting the capillary tube and the hole in the straw wall. To compare the signals going through the straw wall and its end plugs, the right anode segment is equipped with two readout outputs 3 and 4.

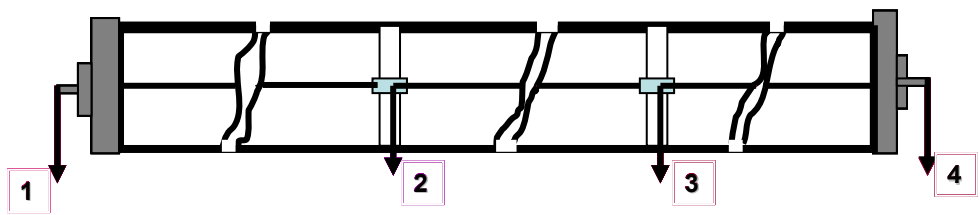

Fig.2 Straw 4mm in diameter contains three anode segments. From the left and middle segments the readout are from the output 1 and 2, correspondingly. From the right segment the readout can be performed either from the output 3 or 4.

When assembling the segmented anodes some capillary tubes are combined with the specialized polycarbonate spacers, forming "assembly units" (Fig.3), without preventing the passage of the gas mixture along the straw. The anodes are installed into the straws under tension of 70g, the contact wires cross some holes in the spacers and are taken out through windows of the sizes ~2x2mm in the corresponding places of the straw walls. After testing over all the readout channels, each window of the straws is gas tightly closed. High voltage can be applied to the straw anodes as well as onto the cathodes in dependence on the readout scheme from the straws.

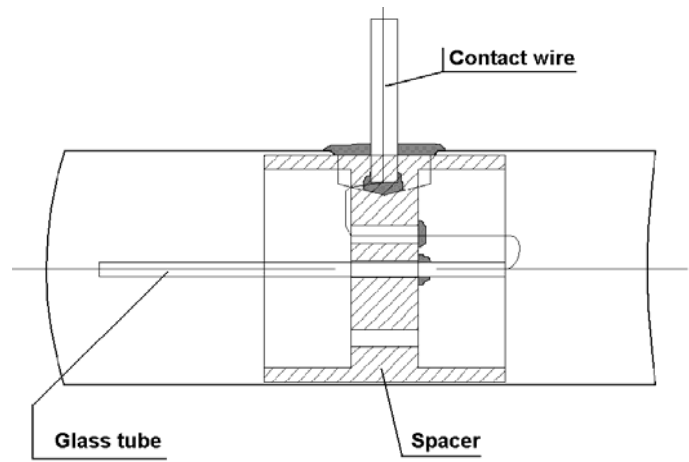

Fig.3. Schematic of the spacer unit in the straw 4mm in diameter.

Several analogous straws shown in Fig. 2, have been assembled and tested. Three anode segments of these straws had a length of ~ 185, 100 and 180mm. The straw under testing was blown through with gas mixture $ArCO_2$ (70/30) and irradiated by the $^{55}Fe$ source through the slit collimator placed perpendicularly to the straw anode. The width of the gamma quaint beam in the median straw plane was ~ 1mm. High voltage was given to the corresponding anode segment. The current amplifier with the input resistance of 300Ω and the sensitivity threshold of 750eV was connected directly to the segment output through the capacitor of 200pf. The test of straw durability to gas has shown the absence of gas leakages. Fig.4 demonstrates the general layout of straw.

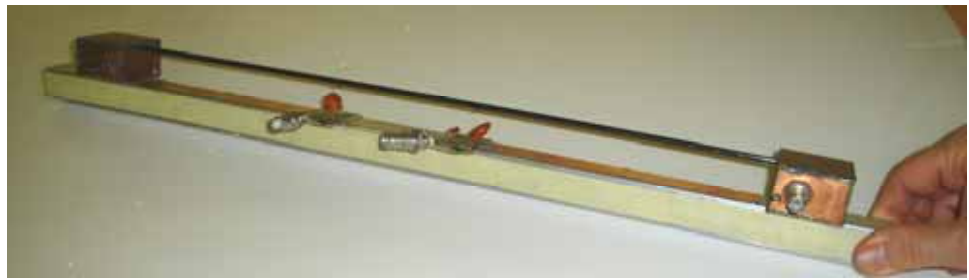

Fig.4. Straw 4mm in diameter and about 465mm long with three anode segments.

Fig.5 shows values of the amplitudes being registered from the corresponding signal segments. The collimator was moved along the straw, for the right segment the signal amplitudes were measured in each position of the collimator from the 3 and 4 outputs by turns. The voltage precision on the anode was better than 1V, the amplitude was measured with one and the same amplifier placed near the outputs 1, 2, 3 and 4 (Fig.2).

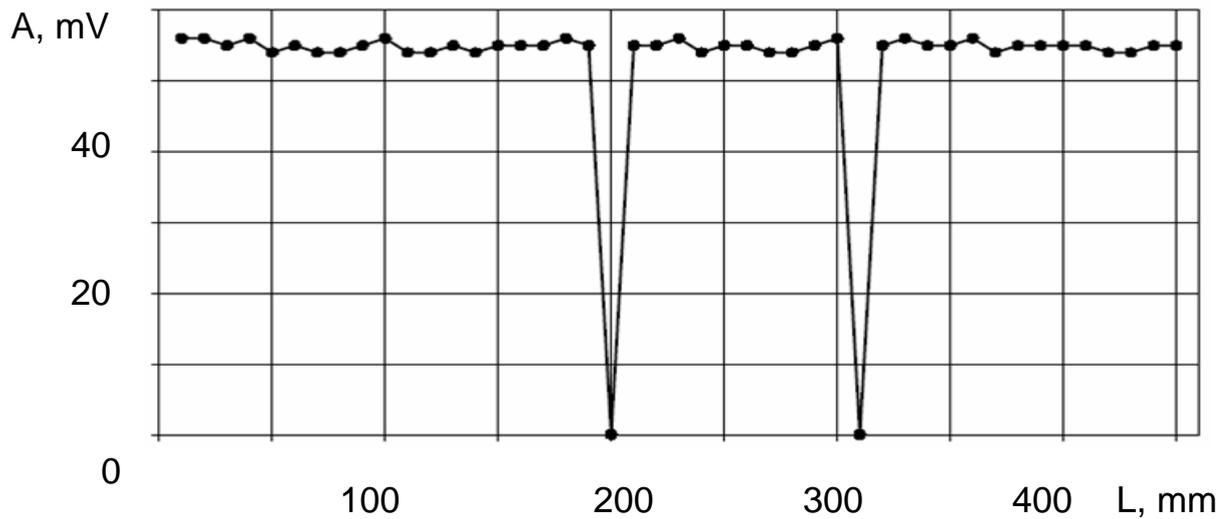

Fig.5. Signal homogeneity along the straw at the fixed voltage and gas mixture.

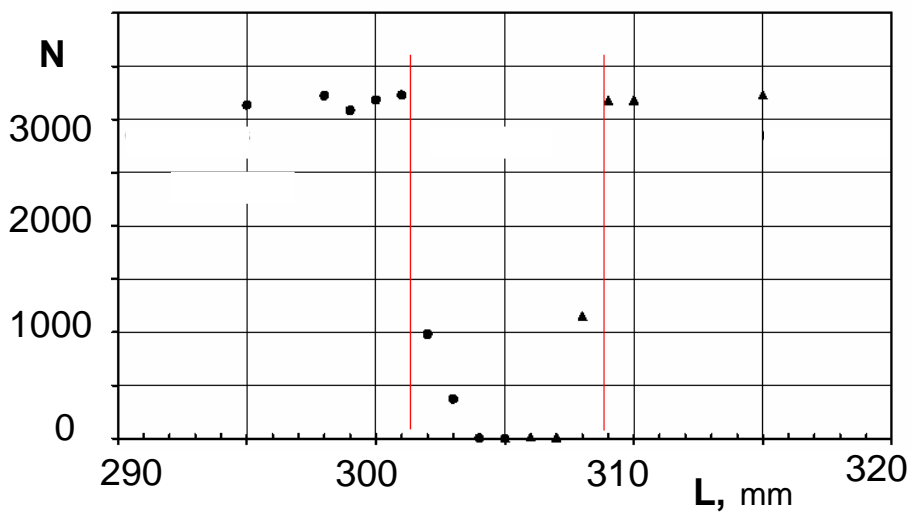

Fig.6. Straw inefficiency on its length in the spacer unit length.

A small slope of the dependence can be explained by the direction of the gas mixture flow. An average deviation of the signal amplitudes D defined with the expression D=2(Amax - Amin)/(Ain + Aout), is not more than 3%. Values Amax, Amin - maximal and minimal signal amplitudes for the straw, and Ain and Aout are the signal amplitudes at the beginning and end of the straw along the direction of the gas mixture flow. Scanning of the spacer units length by using the collimated source, has shown that inefficient straw length is 7.2mm when using a capillary tube 7mm long and a spacer 4.15mm long and 3.95mm in diameter. The radiation thickness of these straw lengths is not more than 0.4%Xo for each. Fig.6 demonstrates inefficiency in the spacer unit area.

Checking of the individual straws with 3 anode sectors have shown their good operation using the gas mixture of $ArCO_2$ in the range of the gas gain till $10^5$, and no discharges were observed between any elements placed under the anode-cathode potentials.

## 3. Detector prototype

To check a real opportunity of manufacturing the straw detectors with a big number of the anode segments, the prototype was made from 19straws 50cm long and 4mm in diameter. The prototype consists of 2 straw layers and neighbouring straws of the layer were glued between themselves. Fig.7. shows the common view of the prototype.

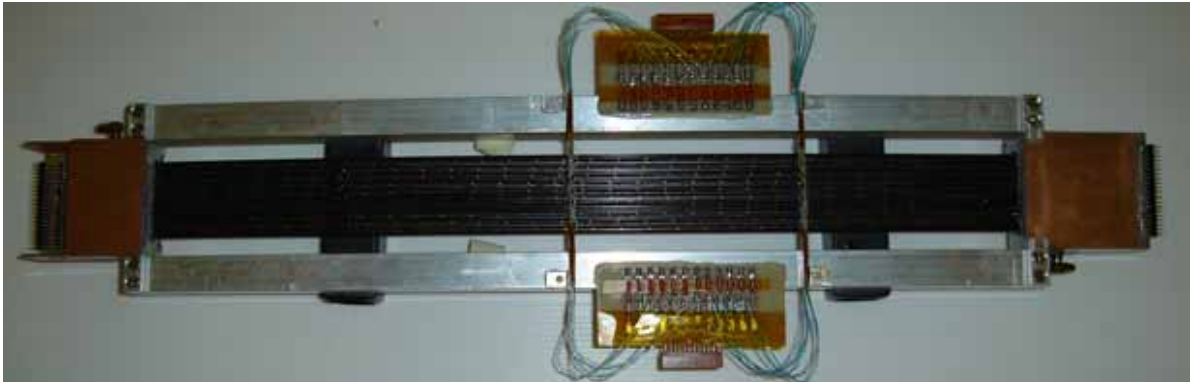

Fig.7. The common view of the prototype with nineteen straws ~500mm long and 4mm in diameter.

Four segmented anodes were installed into 10straws of the top layer (Fig.7), from the nine straws of the second layer the five straws had usual one-segmented anodes and four straws contained three segmented anodes. The prototype with the 19 straws had totally 57 readout channels. Fig.8 shows the electrical scheme of the upper layer straws. The voltage onto the edge anode sectors of the both layers straws is given from the H/V bus via resistors 1M$\Omega$ to the crimping in the end-plugs anode wires, signals through the capacitors are transferred to the connectors 1 and 4 placed nearby. Signals from the middle segments are given to the connections 2 and 3 by using the contact wires of about 15cm long. Resistance of the contact points between the anode and contact wires do not influence the registered signals due to their small values (Fig.9).

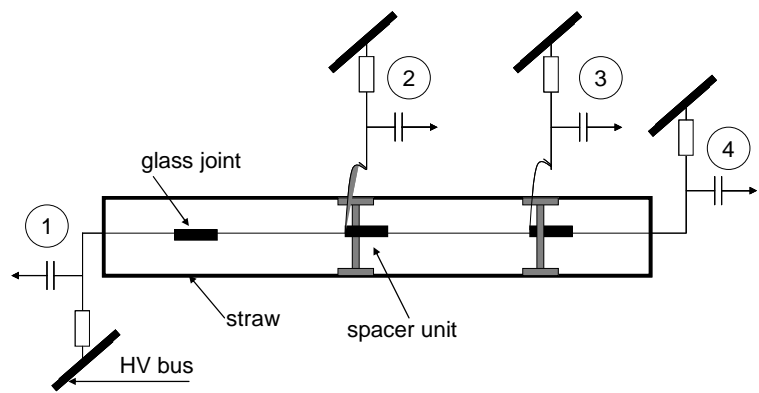

Fig.8. The readout scheme from the straw segments.

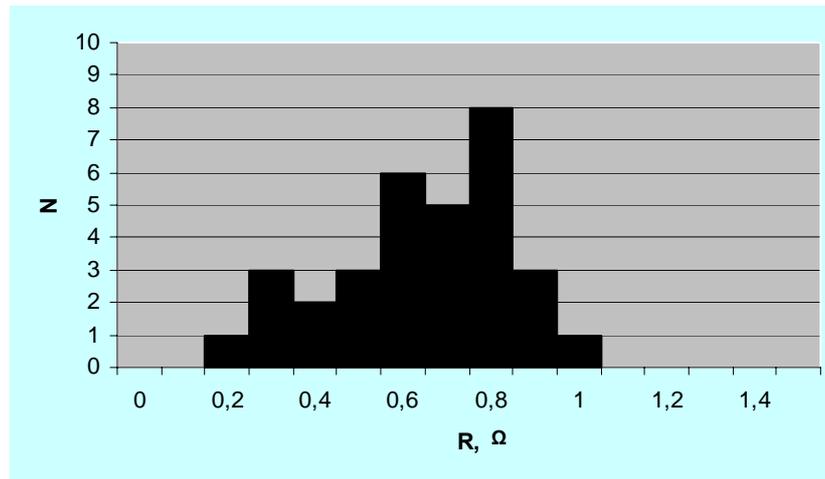

Fig.9. Contact resistance in the capillary tube between the anode and contact wires.

Tests over the prototype over all the registration channels have shown good signal identity at identical parameters of high voltage, environment temperature and partial pressure of the gas mixture (Fig.10).

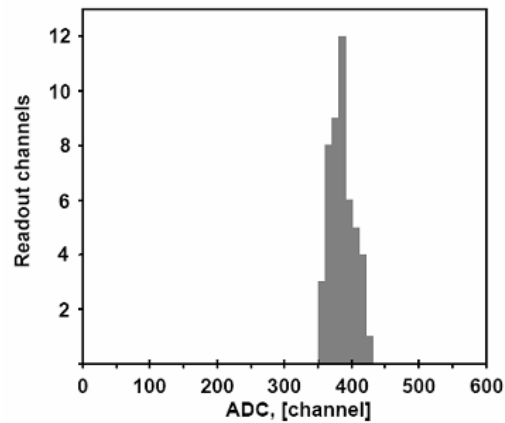

Fig.10. Distribution of the registered signals for the prototype segments. The γ-source is $^{55}$Fe. Gas gain - $2*10^4$.

### 4. Conclusions

The fulfilled studies have shown an opportunity of assembling multi segmented anodes, their installation into the straw with the followed output of the contact wires through the straw walls, the preliminary check of the readout channels with the necessary changing of the working wrong anodes and next gas tight closing the straws. The results are very promising and the study is needed to be continued.